\documentclass[fleqn,usenatbib]{mnras}
\usepackage{newtxtext,newtxmath}
\usepackage[T1]{fontenc}
\usepackage{graphicx}   
\usepackage{amsmath}    

\usepackage{amssymb}    

\title
    [OJ~287: BH mass estimate with BMC]{Black hole mass estimate in OJ~287
     based on the bulk-motion comptonization model}

\author[Sergey Kuznetsov \& Lev Titarchuk]
{Sergey Kuznetsov$^{1}$\thanks{E-mail: astro@prosto.science (SK)}
\&
Lev Titarchuk$^{2}$
\\
$^{1}$Prosto Science AI Lab, Montreal, QC, H2X 3X2, Canada\\
$^{2}$Dipartimentodi Fisica, Universit\`a di Ferrara, via Saragat 1, 44122 Ferrara, Italy
}

\date{Last updated 2024 February 5; in original form 2023 January 24}

\pubyear{2023}

\begin{document}
\label{firstpage}
\pagerange{\pageref{firstpage}--\pageref{lastpage}}
\maketitle

\begin{abstract}
The multi-wavelength outburst activity in the BL Lacertae source
OJ~287 has sparked a lot of controversy about whether the source
contains one or two black holes (BHs) and what characteristics of this
black hole binary would be. In this article we present the results of
analysis of the X-ray flaring activity of OJ~287 using the data of
{\it Swift}/XRT observations. We discovered that the energy spectra in all
spectral states can be adequately fit with the XSPEC BMC model (the
Comptonization one).  As a result we found that the X-ray photon index
of the BMC model, $\Gamma$ correlates with the mass accretion rate,
$\dot M$. We found the photon index $\Gamma$ to increase monotonically
with accretion rate $\dot M$ from $\Gamma\sim 2$ in the intermediate
state (IS) to $\Gamma\sim2.5$ the high/soft state (HSS) with
subsequent saturation at $\Gamma\sim$ 2.6 level at higher
luminosities. This type of behavior of the spectral index is
remarkably similar to the pattern observed in a number of established
stellar-mass black hole candidates. Assuming the universality of the
observed pattern of the correlation between the photon index and mass
accretion rate, we estimate the BH mass in OJ~287 to be around
$2\times10^8$ solar masses, using the well studied BH binaries
GX~339--4 and XTE~J1859--226 to calibrate the model.
\end{abstract}


\section{Introduction}

The primary goal of our paper is to examine the value of the secondary
black hole (BH) mass in OJ~287 applying the scaling method of BH mass
determination [\cite{st09}, hereafter ST09].  OJ~287 demonstrated a
noticeable 12 yr cycle [\cite{val06}] that can be explained by the presence of the 
accreting black hole binary system (BHBS) with systematic disturbance of the 
accretion disk (AD) around the primary by the secondary.

\cite{val12}, hereafter V12, presented results of the two X-ray
observations of OJ~287 by {\it XMM-Newton} on April 12 and
November 3-4, 2005.  In Fig. 2 they plotted the observations of the 2005
campaigns.  V12 claimed that for the April 12, 2005 the spectral energy
distribution (SED), the spectrum from radio to X-rays followed nicely
a synchrotron self-Compton (SSC) model. V12 further stated that the
BH mass value of the secondary determined from the orbit solution
$M_{sec}\sim1.4 \times 10^8$ $ M_\odot$ agrees with the measured
properties of the optical outburst. V12 also formulated a question
regarding the primary BH mass necessary to guarantee the stability of
the primary accretion disc. They found that the minimum value of the
primary mass $1.8 \times 10^{10} M_{\odot}$ is quite close to the BH
mass determined from the orbit, $1.84 \times 10^{10} M_{\odot}$.

To find evidence for the emission of the secondary, V12 needed to look
at the short time-scale variability in OJ~287. It has been found to be
variable from 15 min time-scale upwards on many occasions [see for
example, \cite{gup12}] and on one occasion the light curve has shown
sinusoidal variations with a period of 228 min \citep{sag04}. If this
variation is associated with the last stable orbit of a maximally
rotating black hole, the mass of the black hole would be $1.46\times 10^8
M_{\odot}$ \citep{gup12}, i.e. identical to the mass obtained from the
orbit solution \citep{val10}.

\cite{kom21}, hereafter KOM21 made a detailed analysis of {\it XMM-Newton}
spectra of OJ~287 spread over 15 years. KOM21 also presented data from {\it Swift}
UVOT and XRT observations of OJ~287, which began in 2015, along with
all public {\it Swift} information after 2005. During this period, OJ~287
was found in the low/hard extreme state and outburst soft states. In
addition, they established that the OJ~287 X-ray spectrum was highly
variable, passing all states seen in blazars from the low/hard (flat)
state to exceptionally soft steep state (ST). KOM21 found that these
spectra can consist of two parts: IC radiation which is prevailing
at the low-state, and very soft radiation component that becomes
extremely powerful when OJ~287 becomes more luminous. KOM21 have concluded
that their 2018 {\it XMM-Newton} spectra almost identical with
the EHT examination of OJ~287, and were well characterized by a model with
a hard IC component of the photon index $\Gamma \sim 1.5$ and a soft
component. 

We reanalyzed the data of the X-ray telescope (XRT) onboard of {\it Swift}
Observatory for OJ~287 and applied the ST09 scaling method to this
data in order to make a BH mass estimate. Using this method requires
an accurate knowledge of the distance to the source OJ~287.  The
method was proposed back in 2007 by \cite{st07}, hereafter ST07 and also by
ST09.  It is worth noting that there are two scaling methods: the
first method based on the correlation between the photon index
$\Gamma$ and the quasi-periodic oscillation frequency (QPO) $\nu_L$;
and the second method is based on the correlation between $\Gamma$ and
normalization of the spectrum proportional to $\dot M$. If the
first method ($\Gamma-\nu_L$) is used to estimate the mass of the BH, the
distance to the source is not required (ST07), but for the second  method
($\Gamma-\dot M$) the source distance and the inclination
of the accretion disk relative to the Earth observer are needed
(ST09).

For both methods, it is important that the source shows a change in
spectral states during the outburst and a characteristic behavior of
$\Gamma$, namely, a monotonic increase of $\Gamma$ with $\nu_L$
or $\dot M$ in the LHS$\to$IS$\to$HSS transition and reaching a
constant level (saturating) at high values of $\nu_L$ or $\dot M$,
followed by a monotonic decrease in $\Gamma$ during the HSS$\to$IS$\to$LHS
transition during the outburst decay.

The saturation of $\Gamma$ (so-called ``$\Gamma$-saturation phase'')
during the outburst is a specific signature that this particular object
contains a BH~\citep{tz98}.  Indeed, the $\Gamma$-saturation phase can
be caused only by an accretion flow converging to the event horizon of
a BH [see the Monte-Carlo simulation results in \cite{LT99,LT11}].
Because of this, it makes sense to compare BH sources that have the same
$\Gamma$-saturation levels.  In the second {method} ($\Gamma-\dot M$),
it is assumed that BH luminosity is directly proportional to $\dot M$
(and, consequently, to the mass of the central BH), and inversely
proportional to the squared distance to the source. Thus, {we can
 determine the BH mass by comparing the corresponding track
  $\Gamma-\dot M$ for} a pair of sources with BHs, in which all
parameters are known for one source, and for the other source
parameters except a BH mass are known (for more details on the scaling
method, see ST09).

The scaling method has a number of advantages in determination of a BH
mass  compared to other methods, namely,  the calculation of the X-ray
spectrum originating in the innermost part of disk based on
first-principle (fundamental) physical models, taking into account the
Comptonization of the soft disk photons by hot electrons of the internal
disk part and in a converging flow to BH. 

It is necessary to emphasize that our approach is only 
justifiable if the observed X-ray emission in OJ~287 is indeed produced by
BMC in the accretion flow. In reality, it may be dominated by emission produced 
in relativistic jets, which is a popular explanation of the multi-wavelength 
emission properties of blazars in general. Nevertheless, we apply the scaling 
method to OJ~287 because, as will be shown below, it yields results (specifically, 
the spectral slope vs. normalization dependence) that indicate that at least a 
significant fraction of the observed X-ray emission in this blazar is produced
by the BMC mechanism.

In this paper, based on {\it Swift} data analysis, we estimate a BH
mass in OJ~287 by applying the scaling technique.  In \S 2 we provide
details of our data analysis while in \S 3 we present a description of
the spectral models used for fitting of this data.  In \S
\ref{mass_estimate} we focus on observational results and their
interpretation. In \S \ref{discussion} we discuss the main results of
the paper. In \S \ref{conclusions} we present our final conclusions.

\section{Observation and  Data  Reduction}
\label{data}

We analyzed observational data of OJ~287 carried out by XRT instrument
aboard {\it Swift} observatory from April 2016 to June 2017 (MJD
57501-57918). We processed $\sim$140  ks of pointed observations of
the source. We retrieved the data from the UK {\it Swift} Science Data
Centre Archive\footnote{https://www.swift.ac.uk/archive} and used XRT
data both in Windowed Timing (WT) and Photon Counting (PC)
modes.

The lowest source count rate was detected at the very beginning and at
the end of this observational cycle (April--June 2016, May--June
2017). It corresponds approximately to XRT observations in PC mode
with the flux below $\sim$1~mCrab. This covers 31 PC observations of
$\sim$34 ks in total. The rest of the observations were carried out
during an epoch of strong flaring activity of OJ~287 in the middle of
this period from October 2016 to April 2017 and they consisted of
$\sim$ 105 ks of 109 WT observations. We excluded only 3 very short
observations from our analysis with accumulation time less than 100 s
per ObsID and an additional 6 ones when the source position on the
detector was at the edge of the field of view (FOV).

For our comprehensive analysis we reprocessed original Level~1 XRT data using
standard xrtpipeline screening settings from HEASOFT package v6.30 and
calibration files version 20210915. We also checked PC data for the
pile-up effect and found its possible presence in 2 observations with
contamination radius of 2 detector pixels.

The source spectra were extracted from every individual PC and WT
observations for our preliminary analysis. For the PC mode we used a
30-pixel circular region for the source spectrum and a 60-pixel
circular region taken from source-free area for background
correction. Because of the lack of 2D-image in WT mode, we used a
rectangular 20x40 pixels region co-aligned to the axis of 1D-image for
both source and background spectra.

A typical pointed XRT observation of OJ~287 consists of $\sim$1 ks or
less. As such, a short duration did not allow us to obtain statistically
significant best-fit spectral model parameters. At the same time, the
estimation of the source flux by the applicable model to the spectrum can
be a tool for the further averaging of individual source spectra over
the flux. In order to implement this approach we used the {\tt
tbabs*bmc} (Bulk motion Comptonization) model with fixed
$N_{H}=2.7\times 10^{20}$ and log$(A)=2.0$. The applicable above
absorption toward OJ~287 was adopted from
\cite{willi}\footnote{https://www.swift.ac.uk/analysis/nhtot} and
constant log$(A)$ was used for simplification and reduction of free
model parameters.

The preliminary spectral analysis demonstrated the 0.3--10 keV source
flux variations in the following ranges: 0.5$-$2.6$\times10^{-11}$
and 0.5$-$4.1$\times10^{-11}$ erg~s$^{-1}$~cm$^{-2}$ for PC and WT
observations. Since our goal was the averaging of the source spectra with
similar flux, we grouped 31 PC and 100 WT observations into 7 and 18
flux bins correspondingly.

The procedure of retrieving averaged spectrum is trivial for PC
observations and includes the following steps: reading of re-processed
event files, application of region filters for source and background
spectra extraction with {\tt xselect}; building the accumulated
exposure map with {\tt ximage}; and generation of Ancillary Response
Files (ARFs) with {\tt xrtmkarf}. Similar to the preliminary spectral
analysis described above per individual ObsID, we applied the same
30/60-pixel circular regions for the source extraction and background
correction. 

The retrieving of averaged source spectrum over WT observations is not
trivial if the orientation of the detector does not match among
all observations plus if the source positions on the detector were
shifted. As a result, the averaged 2D-image will be a combination of
1D-images of the same length under different angles and with different
shifts centered to the source position. The hardest task in this
procedure is the proper background extraction and source/background
area correction using event lists and builded exposure maps.

In order to extract the source spectrum from the set of accumulated WT
observations, we used a 20-pixel circular region as a filter with {\tt
xselect}. In this case we extracted the source counts using the same
40-pixel width/length region as we did before for individual WT
observations (see above). For background extraction we used the
annulus centered to the source position with inner/outer radii of
80/120 pixels. Since the detector size is only 200 pixels in width,
we are then getting the same 40 linear pixels of 1D-image on the
detector despite the individual 1D-images orientations and shifts. The
last step in getting of WT averaged spectrum is the manual correction of
the background scaling factor stored in BACKSCAL keyword of the
spectral file. Since the obtained background effective surface is
actually the length and it has the same width of 40-pixels as the source,
we then manually matched this value of BACKSCAL keyword in the header of
the background file to the source spectrum one.

\section{Analysis and Results \label{results}}
\begin{figure}
  \centering
  \includegraphics[width=\columnwidth]{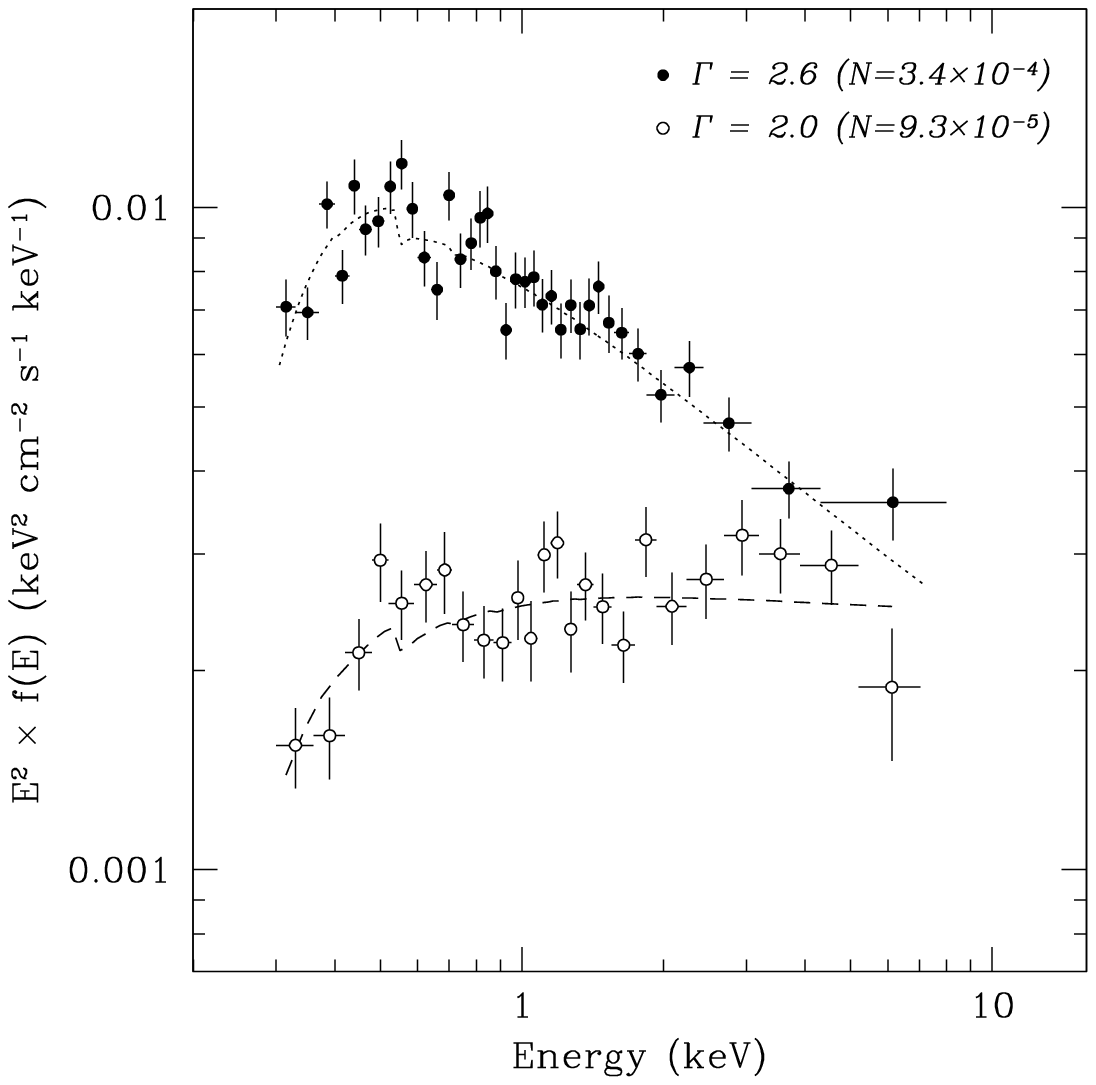}
\caption{
X-ray observed spectra of OJ~287, using the {\tt tbabs*bmc} fit model
for the high/soft (dotted line ) and intermediate (dashed line)
states. Spectra are presented as $E^2 f(E)$-diagram where $f(E)$ is a
photon spectrum. It is evident using this plot that the high/soft
luminosity is at least four times higher than that in the intermediate
state. }
\label{sp_evol}
\end{figure}

For our comprehensive spectral analysis we fitted 7~PC and 18~WT
averaged over the flux spectra by {\tt tbabs*bmc} model with fixed
$N_{H}=2.7\times 10^{20}$ (see above). The typical OJ~287 spectra with
$\Gamma=2.0$ and $\Gamma=2.6$ are shown in Fig.~\ref{sp_evol}.

For our further analysis we plotted the spectral index $\Gamma$ vs $N$ in
Fig.~\ref{three_scal_1}. The correlation between $\Gamma$ and $N$ is
noticeable for $N$ lower than $\sim2.5\times 10^{-4}$. Taking into account that
$\Gamma$ does not fluctuate significantly at higher $N$ values, we
separated our data points shown in Fig.~\ref{three_scal_1} in two
parts and fitted them by linear function and a constant at lower and
higher $N$ correspondingly. The result of this best-fit approximation
is shown in Fig.~\ref{three_scal_1} by a solid line.

Additionally, we fitted the same data points using the analytic function
\begin{equation}
F(N)=A-B\ln\left[ \exp \left(1.0-(N/N_{tr})^\beta \right)+1\right]
\label{eq:params}
\end{equation}
widely used as scalable pattern between reference and target sources
in black hole mass estimation (see ST09). The results of this
approximation are shown in Table~\ref{tab:params}. GX~339-4 best-fit
parameters corresponding to observed LHS to IS source transition in 2007
were taken from ST09 Fig.~3 and Table~2 (28 data points, online
extended table). XTE~J1859-226 data were adopted from ST09 (see Fig.~9 there).

Taking into account $F(N)$ tends asymptotically to a constant at
infinity, we projected the saturation value of $N$ we found for OJ~287 to
$F(N)$. In Fig.~\ref{three_scal_1} it corresponds to the projection of
the break point of the solid curve to the dotted curve. We then 
repeated the same approach in scaling $F(N)$ for GX~339-4 and
XTE~J1859-226. For the reference sources $N$ should be the same way
equidistant from the upper limit of $F(N)$ as $N$ for the target
source. In other words, the relative height of $N$ position between
the asymptotes should be the same.

The curves corresponding to analytical function $F(N)$ of the
reference and target sources from Table~\ref{tab:params} and their
$N_{t}$ and $N_{r}$ are shown in Fig.~\ref{three_scal_2} in details.

\begin{table}
\caption{Parameters of scaling patterns of Eq.~(\ref{eq:params})
  for reference and target sources}
\label{tab:params}
\begin{tabular}{lcccc}
\hline
Source &  $A$ &  $B$ &  $N_{tr}$ & $\beta$ \\
\hline
\hline
XTE~J1859-226 & 2.55 & 1.25  &  7.13E-2  &  1.00 \\ 
GX~339-4      & 2.46 & 0.955 &  1.28E-1  &  1.60 \\
\hline
OJ~287        & 2.54 & 0.42  &  1.33E-4  &  2.28 \\
\hline
\end{tabular}
\end{table}

%
%

\subsection{X-ray spectra\label{sp_analysis}}
To fit the energy spectra of this source we used a {\tt XSPEC} model
consisting of the Comptonization (bulk-motion Comptonization,
hereafter BMC) component [see \cite{tz98, LT99}].  We also used a
multiplicative {\tt tbabs} model \citep{W00} which takes into account
absorption by neutral material.  We assume that accretion onto a BH is
described by two main zones [see, for example, Fig.~1 in \cite{TS21}]:
a geometrically thin accretion disk [e.g. the standard Shakura-Sunyaev
disk, see \cite{ss73} and a transition layer (TL), which is an intermediate
link between the accretion disk, and a converging (bulk) region (see
\cite{tf04}], that is assumed to exist, at least, below 3
Schwarzschild radii, $3R_S = 6GM_{\rm BH}/c^2$.  The spectral model
parameters are the equivalent hydrogen absorption column density
$N_H$; the photon index $\Gamma$; $\log (A)$ is related to the
Comptonized factor $f$ [$={A}/{(1+A)}$]; the color temperature and
normalization of the seed photon blackbody component, $kT_s$ and $N$,
respectively.

Similarly to the ordinary {\tt bbody} XSPEC model,  normalization is a
ratio of the source (disk) luminosity $L$ to the square of the
distance $d$ (ST09, see Eq.~1 there):

\begin{equation}
N=\biggl(\frac{L}{10^{39}\mathrm{erg/s}}\biggr)\biggl(\frac{10\,\mathrm{kpc}}{d}\biggr)^2
\label{bmc_norm}
\end{equation}  

This encompasses an important property of our model. Namely, using
this model one can correctly evaluate normalization of the original
``seed'' component, which is presumably a correct $\dot M$
indicator~\citep{ST11}.  In its turn

%

\begin{figure}
\centering
\includegraphics[width=\columnwidth]{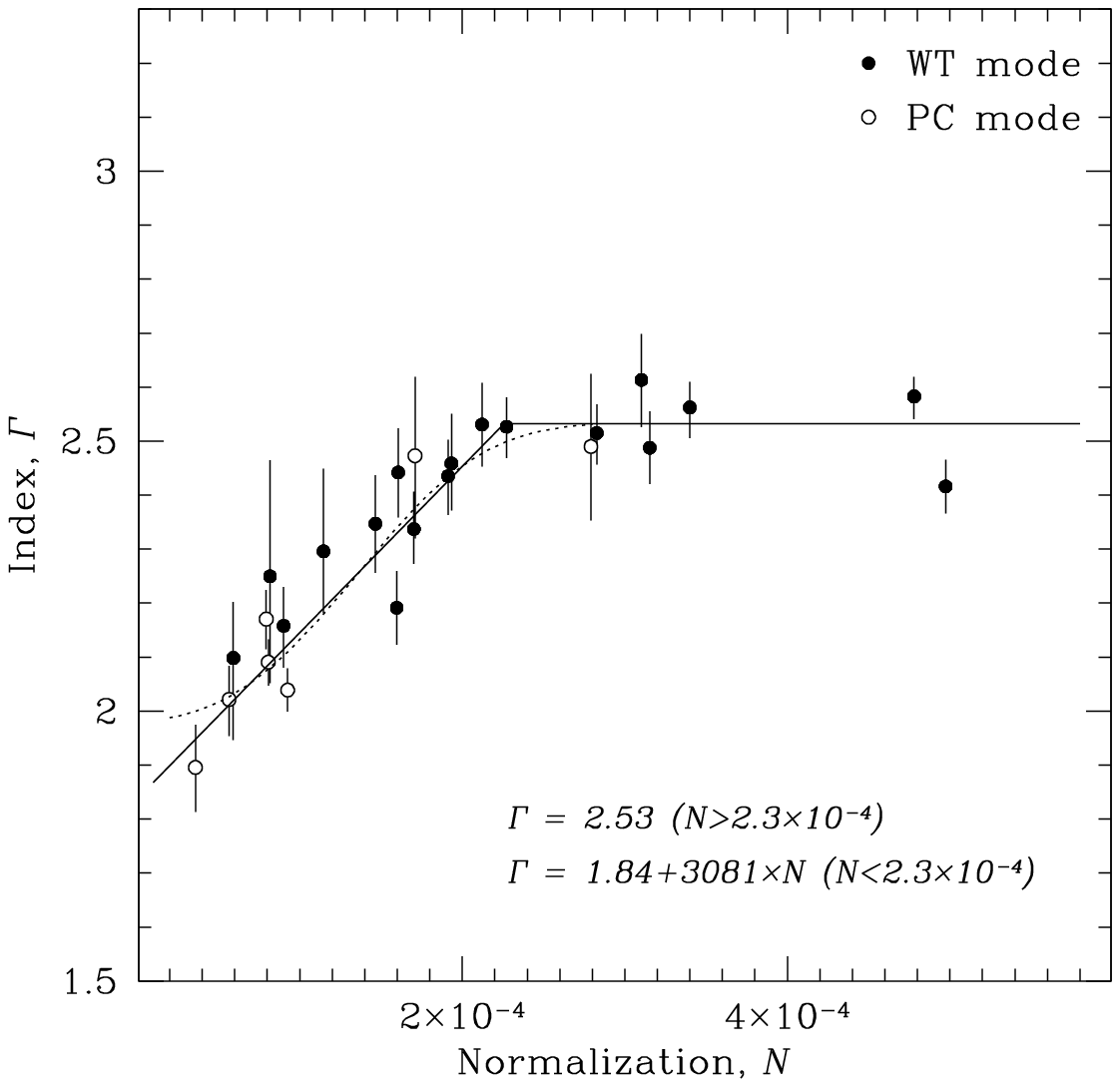}
\caption{ { $\Gamma$ vs $N$ (normalization, proportional to the mass
    accretion rate) correlations those for OJ~287 (target source).  We
    use a specific function for $\Gamma=1.80+3.190\times N$ for
    $N<2.3\times10^{-4}$ and $\Gamma=2.53$ for $N>2.3\times10^{-4}$ to
    fit the data. Dotted curve represents best-fit model shape of Eq~\ref{eq:params}. 
} }
\label{three_scal_1}
\end{figure}

\begin{equation}
L = \frac{GM_{BH}\dot M}{R_{*}}=\eta(r_{*})\dot m L_{Ed}
\label{bmc_norm_lum}
\end{equation}  
Here $R_{*} = r_{*} R_S$ is an effective radius where the main energy
release takes place in the disk, $R_S = 2GM/c^2$ is the Schwarzschild
radius, $\eta = 1/(2r_{*})$, $\dot m = \dot M/\dot M_{crit}$ is the
dimensionless $\dot M$ in units of the critical mass accretion rate
$\dot M_{crit} = L_{Ed}/c^2$, and $L_{Ed}$ is the Eddington
luminosity.

For the formulation of the Comptonization problem, one can see
\cite{tmk97,tz98,LT99,Borozdin99,st09}.

Spectral analysis of the {\it Swift}/XRT data of OJ~287, in principle,
provides a general picture of the spectral evolution.  {We can trace
the change in the spectrum shape during the IS--HSS transition in}
Fig.~\ref{sp_evol}, which demonstrates two representative $E*F_E$
spectral diagrams for different states of OJ~287.  The emergent
spectra (see Fig.~\ref{sp_evol}) can be described as a sum of the low
energy blackbody and its fraction convolved with the Comptonization
Green function (CGF) [see Eqs.~(16) and (B5) in \cite{st80}].  The HSS
and IS spectra are characterized by a strong soft blackbody component
(presumably, related to the accretion disk) and a power law (as the
hard tail of the CGF), while in the LHS the Comptonization component
is dominant and the blackbody component is barely seen because the
innermost part of the disk is fully covered by the scattering media
which has a Thomson optical depth { more than 2}.
  
Analysis of the {\it Swift}/XRT data fits (see Fig.~\ref{sp_evol})
showed that $\Gamma$ monotonically increases from 2
to 2.6, when  normalization of the spectral component (or
$\dot M$) increases by a factor of about 5 (see points, in
Fig. ~\ref{three_scal_1}) at the outburst rise phase (IS--HSS).

%
%

\begin{table*}
\caption{BH masses and distances}
\label{tab:par_scal}
\centering 
\begin{tabular}{llllllc}
\hline\hline                        
Reference sources & $m_r^{(a)}$ (M$_{\odot})$ & $i_r^{(a)}$ (deg) & $d_r^{(b)}$ (kpc) & References  \\
\hline
XTE~1859--226 & 5.4$^{(a)}$ & $\sim70$ & $\sim4.2^{(b)}$ & \cite{corral11},  \cite{hynes02}, \cite{st09}\\  
GX 339-4 &      12$^{(a)}$  &   $\sim70$ &  5.75$\pm0.8^{(a)} $ &  \cite{st09}, \cite{hynes03}, \cite{heida17}\\
\hline\hline                        
Target source   & $m_{t}$ (M$_{\odot}$) & $i_t^{(a)}$ (deg) & $d_t^{(b)}$ (kpc) &  \\
      \hline
OJ~287 & 2$\times10^8$ &   50  &  1.037 Gpc   &   using  XTE~J1859--226 as a reference source \\
OJ~287  & 2$\times10^8$ &  50 &  1.037 Gpc    &   using GX~339-4  as a reference source  
\\
\hline
\\
\end{tabular}
\\ (a) Dynamically and using the scaling method values of BH masses
and system inclination; (b) source distance found in literature.
\end{table*}

%
\begin{figure}
\centering
\includegraphics[width=\columnwidth]{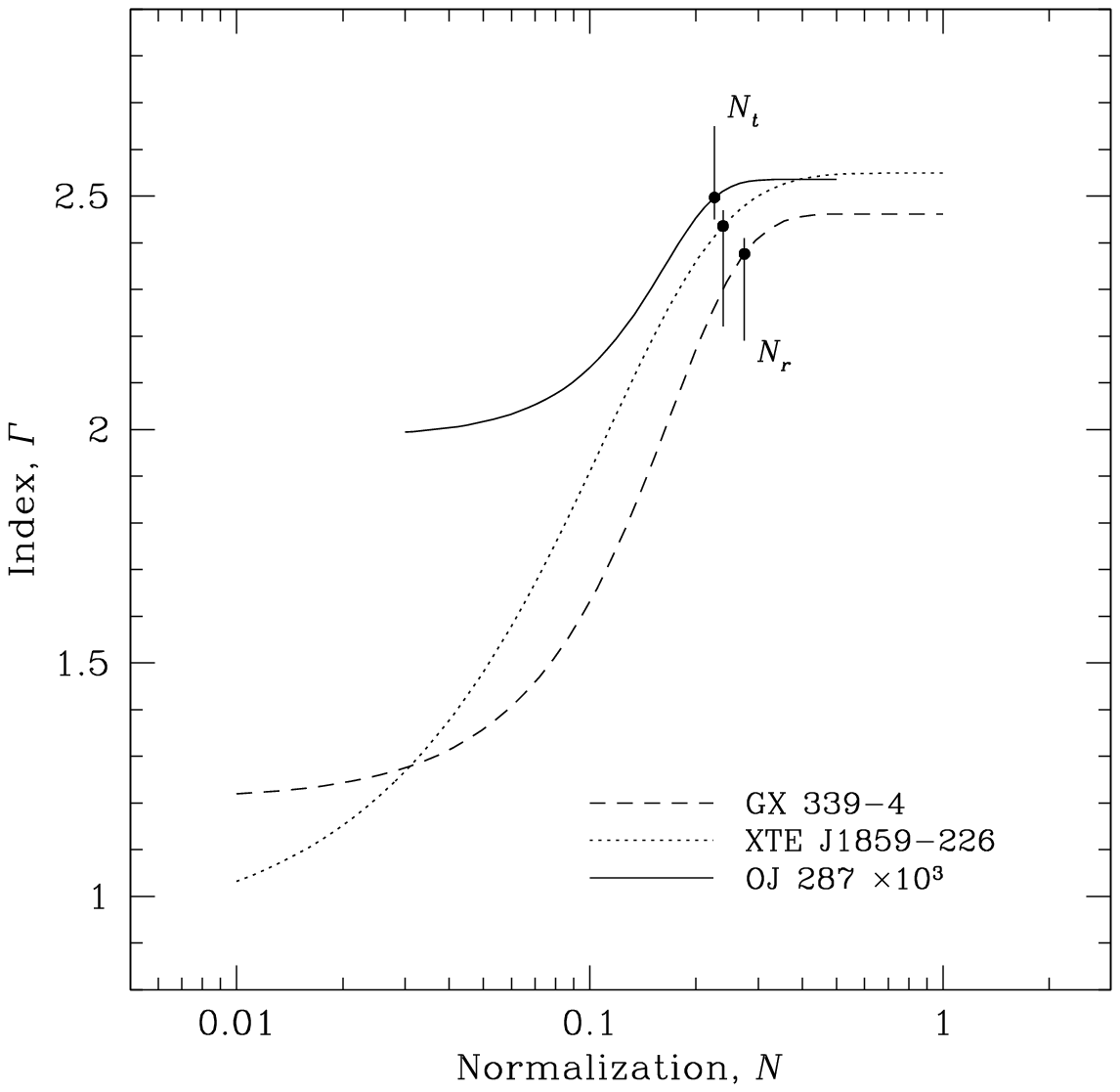}
\caption{ $\Gamma$ vs. N (normalization, proportional mass accretion
  rate and BH mass) for OJ~287 (target source) and GX 339-4 and XTE
  1859-226, the reference sources. The transition from the power line
  (linear shape) to the asymptotic horizontal asymptotic line occurs
  at $N_t=0.23\times10^{-4}$ for OJ~287 and at $N_r=0.32, ~0.31$ for
  GX 339-4 and XTE 1859-226, respectively.  }
\label{three_scal_2}
\end{figure} 

\section{A BH mass estimate}
\label{mass_estimate}

Now, we use the scaling method for a BH mass estimate ($M_{oj287}$) in
OJ~287.  The BH mass scaling method using the $\Gamma-N$ correlation is
described in detail in ST09. This method consists of ({\it i}) searching
for such a pair of BHs for which the $\Gamma$ correlates with
{increasing} normalization $N$ (which is proportional to mass
accretion rate $\dot M$ and a BH mass $M$, see ST09, Eq.  (7) there)
and the saturation level, $\Gamma_{sat}$ are the same and ({\it ii})
calculating the scaling coefficient $s_{N}$ which allows to determine
a BH mass of the target object.  {It is worthwhile to emphasize that
  one needs a ratio of distances for the target and reference sources
  in order to estimate a BH mass using the following equation for the
  scaling coefficient}
\begin{equation}
 s_N=\frac{N_r}{N_t} =  \frac{m_r}{m_t} \frac{d_t^2}{d_r^2}{f_G}
\label{mass}
\end{equation}
{where $N_r$, $N_t$ are normalizations of the spectra,
  $m_t=M_t/M_{\odot}$, $m_r=M_r/M_{\odot}$ are the dimensionless BH
  masses with respect to solar, $d_t$ and $d_r$ are distances to the
  target and reference sources, correspondingly.}  A geometry factor,
$f_G=\cos i_r/\cos i_t$ where $ i_r$ and $ i_r$ are the disk
inclinations for the reference and target sources, respectively [see
  ST09, Eq.~(7)].

We found that XTE~1859--226 and GX 339-4 can be used as the reference
sources because these sources met all aforementioned requirements to
estimate a BH mass of the target source OJ~287 [see items (i) and (ii)
  above].

In Fig.~\ref{sp_evol} we show two types of the spectral states
observed in OJ~287. The first one (in the bottom there) is a typical
intermediate  spectrum observed when the resulting luminosity is relatively
low.  We presented all emergent spectra of this source in terms of
$EF_(E)$ diagrams in order to estimate the resulting
luminosity. In Fig.~\ref{three_scal_1} we demonstrate the
$\Gamma$ versus $N$ where $N$ is presented in the units of
$L_{39}/d^2_{10}$ ($L_{39}$ is the source luminosity in units of
$10^{39}$ erg/s and $d_{10}$ is the distance to the source in units of
10 kpc). As one can see, the correlations of these sources
are characterized by similar shapes and saturation levels
$\Gamma_{sat}>2.5$.  In order to implement the scaling method we use
an analytical approximation $F(N)$ for the $\Gamma-N$ correlation,

\begin{eqnarray}
\Gamma(N) &=& 1.84+3081*N ~~~~ ~~~{\rm for} ~~~~~~~N<2.3\times 10^{-4}\nonumber \\
\Gamma(N) &=& 2.53~~~~~~~~~~~~~~ ~~~~~~~~~~~~{\rm for} ~~~~~~~N>2.3\times 10^{-4} 
\label{scaling function_N0}
\end{eqnarray}

We estimated a BH mass for OJ~287 using Fig~\ref{three_scal_1} of
this manuscript and Figs. 3 and 9 in ST09, for GX 339-4 and XTE1859-226
respectively.  Thus, we found that the $\Gamma$ vs $N$ correlations
are self-similar, at least for $\Gamma>2$, see Fig. \ref{three_scal_2}
and from these correlations we could estimate $N_t$, $N_r$ for OJ~287
and GX 339-4 and XTE1859-226, respectively. They are
$N_t=2.3\times10^{-4}$ and $N_{r, GX}= 0.27$, $N_{r, 1859}= 0.24$.
taken at the beginning of the $\Gamma$-saturation part (see solid
points).

Thus, for GX 339-4 we obtained $s_N=N_r/N_t \sim 0.1\times10^{4}$
applying $N$-values: $N_r=0.27$ and $N_t=2.3\times10^{-4}$ (see
Fig. \ref{three_scal_2}). On the other hand, a value of a factor
$f_G=\cos {i_r}/\cos{i_t} \sim 0.53$ for the target and reference
sources can be obtained using inclinations $i_t=50^{o}$ and
$i_r=70^{o}$ (see Table~\ref{tab:par_scal} and \cite{hynes03}).  As a
result of the estimated target mass (OJ~287), $m_t$ we find that
\begin{equation}
m_t= f_G\frac{m_r}{s_N} \frac{d_t^2}{d_r^2}
\label{mass_target1}
\end{equation}
or
\begin{equation}
m_t\sim
2\times10^8\frac{f_G}{0.53}\frac{m_r}{12}\frac{(d_t/1.073\times10^6)^2}{(d_r/5.75\times
  10^3)^2} (s_N/0.1\times10^4)^{-1}
\label{mass_target1_rev}
\end{equation}
where we use values of $d_t=1.073$ Gpc and $d_r=5.75$ kpc (see Table~\ref{tab:par_scal}).

Also using the reference source XTE J1859-226 we estimate
\begin{equation}
m_t\sim 1.7\times 10^8
\label{mass_target2_rev}
\end{equation}
 applying Eq. (\ref{mass_target1}) if we substitute the values of
 $f_G=\cos i_r/\cos i_t =0.53$, $s_N=0.1\times10^4$, $d_r=4.2$ kpc and
 $m_r=5.4$ (see Table~\ref{tab:par_scal}).

\section{Discussion}\label{discussion}
\cite{val12} suggested that a BH mass of the secondary in OJ~287 is
about $\sim1.4\times 10^{8} M_\odot$ using the orbital solution.
\cite{kom21}, hereafter KOM21 found two X-ray spectral states in OJ~287
(see Fig.4 there).  Their so called 'hard state' dominated by the
power law component and 'soft state' characterized by a soft component
which became very strong when the source was very bright similar to
those presented in Fig. \ref{sp_evol} here.  KOM21 also estimated
the secondary BH mass, $M_{sec}$ as about $1.5\times10^8$ if
$L_x/L_{\rm Edd}< 1.7\times 10^{-2}$. On the other hand KOM21 claimed
that the Eddington ratio $L_x/ L_{\rm Edd}$, that they found,
surprisingly low, $5.6\times 10^{-4}$ with an assumption that
$M_{BH}\sim 1.8\times 10^{10}$ solar masses. In fact, our result
confirmed their guess that $L_x/L_{\rm Edd}$ is on the of order of $2\times
10^{-2}$ and thus, a BH mass is about $2\times10^8$ solar masses and
is related to the secondary BH in the OJ~287.

Our analysis of the Fall 2016 -- Spring 2017 Swift observations
identified two distinct spectral states of OJ 287: the intermediate
and a soft one. Those states correspond to the initial rise of the X-ray
spectral index correlated to increase of the mass accretion rate
(intermediate state) followed by saturation of the spectral index
after reaching a certain level of accretion rate. The results of
spectral analysis of the intermediate and soft states were used to
estimate the value of the mass accretion rate corresponding to the
saturation of the spectral index and consequently constrain the BH
mass. The third spectral state of OJ~287 that can be characterized by
harder source spectrum and represents a different accretion regime,
was not the subject of our analysis.

\section{Conclusions}\label{conclusions}
The multi-wavelength outburst activity in OJ~287 with X-ray telescope
(XRT) on the {\it Swift} call into question whether the source contains
one or two black holes (BHs). It is very important to reveal the
characteristics of this binary. In the presented work we demonstrated
that the OJ~287 X-ray spectra underwent the state transition from the
intermediate state (IS) to the high/soft state (HSS) (see
Fig. \ref{sp_evol}).  We found that energy spectra in all spectral
states can be modeled using a product of the {\tt tbabs} and a Comptonization
component, BMC (see XSPEC).  Moreover,  in OJ~287 we discovered the
correlation of the photon index $\Gamma$ with normalization, N (proportional
to the disk mass accretion rate $\dot M$, see Fig. \ref{three_scal_1})
similar to those established in BH Galactic sources by ST09.  We found
that $\Gamma$ increases monotonically with $\dot M$ from the
intermediate state (IS) to the high-soft state HSS, and then saturates
at $\Gamma\sim$ 2.6.  This can be considered as observational
evidence of the presence of a BH in OJ~287. Based on this correlation,
we apply the scaling method (ST09) to estimate a BH mass
$\sim2\times 10^8$ 
solar masses, using the
well-studied X-ray BH binaries, GX 339--4 and XTE J1859--226 as
reference sources (see Fig. \ref{three_scal_2}).

\section*{Acknowledgements}
This work made use of data supplied by the UK {\it Swift} Science Data
Centre at the University of Leicester. The authors are grateful to Dr.
Kim Page of the UK {\it Swift} Science Data Centre for the help with
{\it Swift}/XRT data processing and analysis. 

The authors would like to thank Sergey Sazonov, Alexey Vikhlinin and
Sergey Trudolyubov for careful reading of the manuscript and valuable
suggestions.  We are especially grateful to Colm Elliott, Marla
Kouri-Towe and Michele Delmaire for the proof reading and useful
suggestions about corrections to the text of the manuscript.

\section*{Data availability} 
The {\it Swift} data is available to download through the UK Swift Data Science
website \url{https://www.swift.ac.uk/archive}. 

\bibliographystyle{mnras}

\label{lastpage}
\end{document}